\documentstyle[aps,preprint]{revtex}
\begin{document}

\title{A toy model of Polyakov duality}
\author{Vipul Periwal}
\address{Department of Physics,
Princeton University,
Princeton, New Jersey 08544}
\date{PUPT-1889}
\def\DD{\hbox{D}}
\def\dd{\hbox{d}}
\def\tr{\hbox{tr}}\def\Tr{\hbox{Tr}}
\def\ee#1{{\rm e}^{{#1}}}
\def\part{\partial}
\def\ln{\hbox{ln}}
\def\cz{{\cal Z}}
\def\bpart{\bar\partial}
\def\del#1#2{{{\delta #1}\over{\delta #2}}}
\def\refe#1{eq.~\ref{#1}}
\maketitle\tightenlines
\begin{abstract}
Polyakov has conjectured that Yang--Mills theory should be equivalent 
to a noncritical string theory.   I pointed out, based on the work of  
Marchesini, Ishibashi, Kawai and collaborators,
and Jevicki and Rodrigues, that the loop operator of the 
Yang--Mills theory is 
the temporal gauge string field theory Hamiltonian of a
noncritical string theory.  In the present note I 
explicitly show how this works for the one--plaquette model, providing 
a consistent direct string interpretation of the unitary matrix model 
for the first time.  
\end{abstract}
\bigskip
The naturality of a string interpretation of Yang--Mills theory 
was first appreciated by Mandelstam\cite{sm}, and has been elaborated 
upon with significant insights by many others\cite{others}.  The 
Wilson loop observables of gauge theories satisfy dynamical 
Schwinger--Dyson equations that have precise geometric 
interpretations\cite{loop}.  These loop equations simplify 
considerably in the limit $N\uparrow \infty$ where $N$ is the
rank of the gauge group.  This is also natural for the string 
interpretation following `t Hooft's study of the large $N$ limit of
gauge theories, since this is the limit when string loop 
contributions to amplitudes are suppressed.  Thus Wilson loop 
expectation values in 
the large $N$ gauge theory satisfy classical equations.  
The loop equation appears also as the saddle--point equation in the 
collective field theory setup of Jevicki and Sakita\cite{coll}.

Strings do 
not interact in this limit---yet the loop equation has terms 
corresponding to the splitting of strings, and terms that annihilate 
strings.  This, of course, is completely consistent, but not entirely 
obvious to the string theorist steeped in the lore of critical string 
theories in conformal gauge.  Indeed, the simplest physical 
interpretation of the loop equation and its meaning in a string 
theory equivalent to the Yang--Mills theory is obtained by 
combining\cite{me}\ two beautiful results.  The first of these is 
the observation of Marchesini\cite{marc}\ that the Fokker--Planck 
Hamiltonian that arises in Parisi--Wustochastic quantization\cite{pw}\  
is precisely the loop operator, {\it i.e.} the operator $H$ such that
\begin{equation}
\langle H\prod_{i}W(C_{i}) \rangle = 0 
\end{equation}
are the Schwinger--Dyson loop equations.  
The second result is that of Jevicki and 
Rodrigues\cite{jev2}: the temporal gauge noncritical string field 
theory Hamiltonian found in the work of 
Ishibashi, Kawai and collaborators\cite{ikall}\ is the Fokker--Planck 
Hamiltonian of the matrix model representation of the noncritical 
string theory. (For earlier relevant work see \cite{das,moore}.)
Combining these results, I found that the 
loop operator of the Yang--Mills theory can be interpreted as the 
exact
string field theory Hamiltonian of a noncritical string theory, {\it 
provided} a crucial consistency condition is satisfied: the loops 
must be defined in a manner that preserves the zig--zag symmetry 
emphasized by Polyakov\cite{pol}. 
The string theory defined by the loop operator is valid at arbitrary 
$N.$  

Now loop equations are notorious for requiring careful 
regularization\cite{dot}, so one must make such formal statements precise with
well--defined operators.  Fortunately, since one is automatically in 
the framework of stochastic quantization\cite{pw,stochrev}\ there is in
fact a continuum regularization available\cite{halpern}.  I derive in 
detail the
noncritical string theories one obtains in this manner in 
\cite{ymsft}, but the aim of the present paper is to present a simple 
toy model in which all the features mentioned in the previous 
paragraph are as transparent as can be.  In the process, one finds a
resolution for the (decidedly minor) puzzle of a string interpretation of the 
unitary matrix model to which I now turn.

The depth of the large $N$ limit is
nowhere more apparent than in the model of one Hermitian matrix first 
solved by Br\'ezin, Itzykson, Parisi and Zuber\cite{bipz}.  The 
analogue of their model for gauge theories is the one--plaquette model
solved by Gross and Witten\cite{gw}.  This model is defined by a Wilson
lattice gauge theory action but involving only one plaquette:
\begin{equation}
\cz = \int \dd U \ \exp\left(\beta \tr \left[V(U) + 
V(U^\dagger)\right]\right)
\end{equation} 
where $U$ is a unitary matrix in U$(N).$
This model is solved as follows:  $U$ can be diagonalized since the 
action and the measure are invariant under $U\rightarrow \eta U\eta^{\dagger}.$ 
There is a Jacobian that appears in this process
so we end up with
\begin{equation}
\cz \propto \int \prod\dd\theta_{i} |\Delta(\theta_{i})|^{2}
\exp\left(\beta \sum \left[V(\ee{i\theta_{j}}) + 
V(\ee{-i\theta_{j}})\right]\right)
\end{equation}
where $\Delta$ is the Vandermonde determinant 
$\prod_{j>i}\left(\ee{i\theta_{i}}-\ee{i\theta_{j}}\right).$
This integral can be evaluated by saddlepoints in the large $N$ limit 
when $N/\beta \equiv \lambda$ is held fixed.  Gross and 
Witten\cite{gw}\ found a third--order phase transition in this 
model.  This transition occurs when the coupling  $\lambda$ is strong enough 
that the eigenvalue distribution has support on the entire unit circle 
in the complex plane.  

The  double--scaling limit\cite{double}\ solves the Hermitian matrix 
model\cite{bipz}\ to all orders in $1/N,$ and the random surface 
interpretation dual to the Feynman diagrams of the perturbative 
evaluation\cite{biz} 
of the model leads to a connection with quantum gravity in two 
dimensions\cite{kpz}.  The double--scaling limit can also be used to 
solve the one--plaquette model in a monkey--see, monkey--do\cite{ps}\ fashion
but there is no convincing random surface interpretation of this model.
The best attempt is presented in \cite{djmw}, but this is not directly 
on the basis of observables in the matrix integral, but rather in 
terms of transformations of integrable hierarchy equations that arise 
in these models.  

This lack of an interpretation (see, for example, \cite{gn}) is 
surprising.  In fact, it seems to be entirely due to an
elision in the literature of the fact that 
this model is a gauge theory on a single plaquette.  The natural 
observables in such a gauge theory are   $\Tr U^{n} \equiv \tr 
U^{n}/N,$ where $n$ takes positive or negative integer values.  These 
are just Wilson loops corresponding to paths wound around the 
plaquette $n$ times, {\it i.e.} in terms of the discussion at the 
beginning of this paper, {\it strings} wound around the plaquette.

Let us derive the Schwinger--Dyson equations for this model.
By the right translation invariance of the unitary group Haar measure, 
we have
\begin{equation}
  \int \dd U \ \ee{ \beta {\rm tr} \left[V(U) + 
V(U^\dagger)\right]} \Tr\left(T_{\alpha}U^{n}\right)
= \int \dd U \ \ee{\beta {\rm tr} \left[V(U\ee{H}) + 
V(\ee{{-H}}U^\dagger)\right]}
\Tr\left(T_{\alpha}(U\ee{H})^{n}\right)
\end{equation}
where $T_{\alpha}$ is a normalized anti--Hermitian matrix and $H\equiv 
\sum_{\beta}H^{\beta}T_{\beta}$ is an arbitrary anti--Hermitian matrix 
expanded in terms of the $N^{2}$ basis matrices  $T_{\beta}.$
Using  the identities
\begin{equation}
\sum_{\alpha} \tr AT_{\alpha}BT_{\alpha} = -\tr A\tr B,\qquad 
\sum_{\alpha} \tr AT_{\alpha}\tr BT_{\alpha} =-\tr AB\ 
\label{ident} 
\end{equation}
after differentiating with respect to $H^{\alpha}$ and summing over 
$\alpha,$ we find\cite{rod} (in the large $N$ limit, for the simplest case 
$V(U)=U$)
\begin{equation}
{1\over \lambda} \left(z_{{n+1}}-z_{n-1}\right) 
+\sum_{p=1}^{n}z_{p}z_{n-p} = 0, \qquad n\ge 1.
\label{loopeq}
\end{equation}
Here we have defined $\langle \Tr U^{n}\rangle \equiv z_{n},$ so in 
particular $z_{0}=1.$  This \refe{loopeq}\ is precisely the loop 
equation\cite{loop}\ 
in this model.  This recursion relation can be used to 
determine all the observables $z_{n}$ as a function of $\lambda,$ but
it should be noted that this equation does not fix $z_{1}.$  For 
general $V,$ we would need to evaluate $2d-1$ $z_{i}$ independently 
where $d$ is the degree of the polynomial $V.$  Rodrigues\cite{rod}\ 
studied   \refe{loopeq} in detail. 
In the present case, we shall see that it is 
possible to fix $z_{1}$ by finding the Fokker--Planck Hamiltonian
underlying \refe{loopeq}, and demanding that the amplitudes computed 
from this Hamiltonian reproduce \refe{loopeq}.
 
The loop operator derived from \refe{loopeq}\ is just (for $N=\infty$)
\begin{equation}
 H_{\infty}\equiv T(\lambda) a_{1} +
 \sum_{n>0}n\bigg[ {1\over\lambda}
 \big(a^{\dagger}_{n+1}-(1-\delta_{n,1})a^{\dagger}_{n-1}\big) 
 +a^{\dagger}_{n}+\sum_{p=1}^{n-1} 
 a^{\dagger}_{p}a^{\dagger}_{n-p}\bigg] a_{n} + \hbox{$n<0$ terms}
\end{equation}
where I have introduced creation and annihilation operators 
$[a_{n},a^{\dagger}_{m}]=\delta_{{n,m}},$  for 
$ |n|>0,$ which create and annihilate strings of 
winding number $n.$ In the absence of string merging, 
strings with positive or negative winding number do not interact.

The string propagates in stochastic time, identified on the basis of 
\cite{ikall,jev2,ikn}\ with Liouville time in a noncritical string 
theory.  
Physical correlation functions are obtained, 
as standard in stochastic quantization\cite{pw,stochrev}, via 
\begin{equation}
\lim_{\tau\uparrow\infty}\ \langle 0| \ee{-\tau 
H_{ \infty}}\prod_{i}a^{\dagger}_{i}|0\rangle \equiv \prod_{i}z_{i}
\end{equation}
in the large $N$  factorization since the strings cannot merge in 
this limit.
We need to fix $T(\lambda).$  We require therefore 
\begin{equation}
\lim_{\tau\uparrow\infty}\ \langle 0| \ee{-\tau 
H_{ \infty}}a^{\dagger}_{1}|0\rangle = z_{1}\ .
\label{zone}
\end{equation}
Following \cite{ikn}, the existence of the large $\tau$ limit and
$H_{\infty}|0\rangle=0$ implies
\begin{equation}
\lim_{\tau\uparrow\infty}\ \langle 0| \ee{-\tau 
H_{ \infty}}[H_{ \infty},a^{\dagger}_{1}]|0\rangle = 0 .
\label{wonder}
\end{equation} 
Since a string of winding number 1 cannot split, \refe{wonder}\ gives
\begin{equation}
T + z_{1} + {1\over \lambda} z_{2} =0.
\label{TT}
\end{equation}
Comparing \refe{TT}\ to \refe{loopeq}\ we find
$T = -1/\lambda.$  A simpler way to write $H_{\infty}$ is therefore
\begin{equation}
H_{\infty}=
 \sum_{n>0}n\bigg[ {1\over\lambda}
 \big(a^{\dagger}_{n+1}-(1-\delta_{n,1})a^{\dagger}_{n-1} - 
 \delta_{n,1}\big) 
 +a^{\dagger}_{n}+\sum_{p=1}^{n-1}
 a^{\dagger}_{p}a^{\dagger}_{n-p}\bigg] a_{n} + \hbox{$n<0$ terms}\  
\end{equation}
which could   have been anticipated.
Consider the physics of the two limiting values of $\lambda.$  For small
$\lambda$ the kinetic term dominates and the string grows or shrinks 
but does not split very often---the dominant splitting is into 
one big string and one small string that is then annihilated.  
For large $\lambda$ the string splits all the 
time, and the tadpole is small as well.  It is easy to see that in 
this limit $z_{1}=0,$ which is of course obvious from $\cz.$  In the 
eigenvalue picture, the eigenvalue distribution is concentrated in a 
sector on the circle in the weak coupling ($\lambda$ small) limit 
whereas it spreads over the whole circle at strong coupling.  
Presumably the phase transition occurs when string splitting is 
compensated by string growth so that the propagation of the string has 
a self--similar character.
 
Having found $T$  in this model, we 
can evaluate $z_{1}$ just using $H_{\infty} $ in \refe{zone} as 
claimed.  
In this sense, the 
one--plaquette model is complete  
since we used \refe{loopeq}\ to derive the form of 
$H_{\infty}$ anyway.
(It is of course possible to  evaluate $z_{1}$ independently.
First of all, notice that on account of the symmetry of the measure 
and the action under inverting $U,$ we have $z_{i}=z_{{-i}}.$
Next, we obviously have
\begin{equation}
z_{1}+z_{-1} =  {1\over N }{\part\over\part\beta}\ln\cz
\end{equation}
and $\cz$ is easily evaluated following \cite{gw}.)  

Consider now finite values of $N.$  Correlations no longer factorize 
so we  consider more general Schwinger--Dyson equations:
\begin{equation}
\sum_{\alpha} {\part\over{\part H^{\alpha}}} 
\int \dd U \ \ee{ \beta {\rm tr} \left[V(U\ee{H}) + 
V(\ee{{-H}}U^\dagger)\right] }\sum_{j=1}^{m}n_{j}
\Tr\left(T_{\alpha}(U\ee{H})^{n_{j}}\right)
\prod_{i\not=j}\Tr\left((U\ee{H})^{n_{i}}\right)\biggm|_{H=0}
= 0
\end{equation}
$n_{i}$ need no longer be all positive or all negative, since we must
allow for strings that wind in either direction.  We now find a new 
term, involving the joining of strings:
\begin{equation}
\bigg\langle \sum_{j} (H_{\infty}W(n_{j}))\prod_{i\not=j}W(n_{i}) +
{1\over N^{2}} \sum_{j}\sum_{i\not=j}n_{i}n_{j} W(n_{i}+n_{j})
\prod_{k\not=i,j}W(n_{k}) \bigg\rangle= 0
\label{finite}
\end{equation}
where we have defined $\Tr U^{n_{i}}\equiv W(n_{i}).$  The extra 
factor of $N^{-2}$ arises   because the traces in 
\refe{ident}\ are not normalized.  While we can no longer write a 
recursion relation for the single operator expectation values $z_{i},$ 
we can write down the 
finite $N$ Hamiltonian from \refe{finite}:
\begin{equation}
H_{N} = H_{\infty} + {1\over  N^{2}} 
\sum_{i,j}  ija^{\dagger}_{ {i}+ {j}}a_{i}a_{j},
\label{Nham}
\end{equation}
implicitly replacing $a^{\dagger}_{0}\rightarrow 1.$
A striking difference between this Hamiltonian (\refe{Nham}) and  the 
Hamiltonian in \cite{ikn}\ is that here we have strings with 
positive and negative windings, whereas in \cite{ikn}\ the string 
length alone is the string field parameter.  Recall that in the case 
of orthogonal polynomials on a circle it is a similar appearance of 
negative and positive powers of $z\equiv \ee{i\theta}$ that results in
distinct recursion relations compared to orthogonal polynomials on the 
real line.  It is important to note here that the string joining term
is to be treated in perturbation theory for consistency.

Given the Hamiltonian $H_{N}$, we can write down a string field theory.
Written in holomorphic co\"ordinates, the functional integral for the 
stochastic time evolution is
\begin{equation}
U(a_{i}^{\ast},a_{j};\tau_{f},\tau_{i}) =\int 
\prod_{n,\tau} {{\dd a_{n}^{\ast}\dd a_{n}}\over
{2\pi i}}\ee{{1\over2}\sum_{n}[a_{n}^{\ast}(\tau_{f})a_{n}(\tau_{f})+
a_{n}^{\ast}(\tau_{i})a_{n}(\tau_{i})]}\ee{-\int^{\tau_{f}}_{\tau_{i}}L_{FP}}
\label{funct}
\end{equation}
with the Lagrangian
$L_{{FP}} = {(1/2i)}\sum_{n\not=0}\left(\dot a_{n}^{\ast}a_{n}-
a_{n}^{\ast}\dot a_{n}\right) - H_{FP}$
and
\begin{equation} 
  H_{FP}\equiv\sum n\bigg[{1\over\lambda}\big(a^{\ast}_{n+1}- a^{\ast}_{n-1}  
  \big)  +a^{\ast}_{n}+\sum_{p=1}^{n-1}
 a^{\ast}_{p}a^{\ast}_{n-p}\bigg] a_{n} + \hbox{$n<0$ terms}+ {1\over  N^{2}} 
\sum_{i,j}  ija^{\ast}_{ {i}+ {j}}a_{i}a_{j}  \ .
\end{equation}
(Recall $a^{\ast}_{0}\equiv 1.$)
For our purposes, we require vacuum boundary conditions at 
$\tau_{f}=\infty,$ and the specified Wilson loops in the correlation 
function at $\tau_{i}=0.$  In nontrivial models\cite{ymsft}\ the 
regularization needed can only be implemented at the level of the 
loop operator, not in the functional integral derived from the loop
operator\cite{halpern}.  Another striking feature of the functional 
integral, not perhaps most obvious in \refe{funct}, is the appearance 
of supersymmetry in the string field theory as usual in stochastic 
quantization\cite{gozzi,stochrev}.

There is much more that one can derive explicitly in these models
especially taking finite $N$ corrections into account.  
The explicit solution found in \cite{ps}\ can be used as a check on 
this formalism. The main point of 
this paper however is that there is a string interpretation of the 
simplest Yang--Mills theory, the one--plaquette model.  This string 
interpretation is precisely as stated in  \cite{me}, as a 
noncritical string field theory in temporal gauge, and therefore is 
a concrete example of Polyakov's conjecture\cite{pol}.  The zig--zag 
symmetry is of course built into the formalism here, but needs much 
more work to maintain in nontrivial models\cite{ymsft}.   
 
Turning to the interpretation of \cite{djmw}, it could be that the
mapping between the unitary matrix model and models of closed and open 
strings found in \cite{djmw}\ is hiding some deeper duality between
closed string and open and closed string theories.  The Polyakov 
duality between Yang--Mills theory and a {\it noncritical} 
closed string theory is what has been demonstrated here in a toy model.

I am grateful to L. Yaffe for asking the question this note answers, 
and especially  to G. Lifschytz for sharing his insights. 
I acknowledge a helpful conversation with I. Klebanov and   helpful 
comments by A. Polyakov.  
This work was supported in part by NSF grant PHY-9802484.
\def\np#1#2#3{Nucl. Phys. B#1,  #3 (#2)}
\def\prd#1#2#3{Phys. Rev. D#1, #3 (#2)}
\def\prl#1#2#3{Phys. Rev. Lett. #1, #3 (#2)}
\def\pl#1#2#3{Phys. Lett. B#1, #3 (#2)}

\end{document}